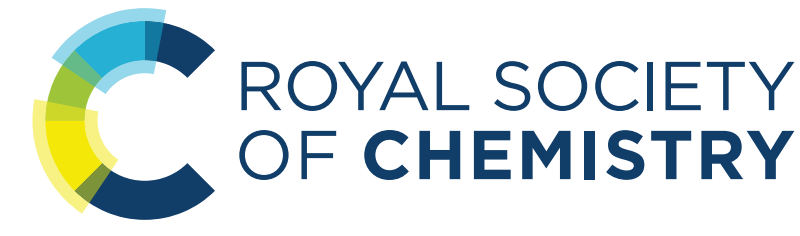

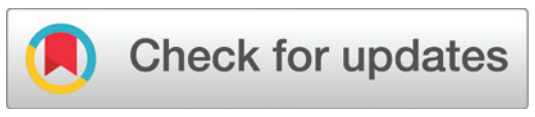

# Hydroflux crystal growth of alkali tellurate oxide-hydroxides

Madalyn R. Gragg, †[a] Allana G. Iwanicki, *†[b,c] Maxime A. Siegler [b] and Tyrel M. McQueen [b,c,d]

This study investigates the synthesis of novel magnetic materials *via* hydroflux synthesis, a method that combines flux-based and hydrothermal techniques. Single crystals of three novel alkali tellurate oxide-hydroxides were synthesized. One, $CsTeO_3(OH)$, is nonmagnetic and a new member of the series $ATeO_3(OH)$ (A = alkali). The other two phases contain magnetic Cu–Te substructures, one of which, $KCu_2Te_3O_8(OH)$, is structurally three-dimensional and undergoes several magnetic ordering transitions. The other, $Cs_2Cu_3Te_2O_{10}$, is structurally two-dimensional and remains paramagnetic above $T$ = 2 K. These exploratory investigations of novel phase spaces reveal key factors, including hydroxide concentration, precursor solubility, and oxidizing power of the solution, in the formation and composition of alkali tellurate oxide-hydroxides.

## I. Introduction

In recent years, complex fluxes have been revived as a method to explore new phase spaces as their solution properties are distinct from individual flux components.[1–4] The hydroflux, one such complex flux combining $H_2O$ and alkali hydroxide (AOH, A = alkali), creates a reaction environment (and also reagent) distinct from either water or alkali hydroxide individually.[5,6] Like many fluxes, the hydroflux enables the formation of metastable phases, products that are formed in non-global energy minima, at lower temperatures ($T \approx 180$–250 °C (ref. 2, 7 and 8)). This is due to the increased diffusion and role of kinetics over thermodynamics. Low temperature metastable phases can contain unusual bonding geometries[9,10] which can be conducive for emergent properties including novel magnetic exchange and superconductivity.[11–15] The hydroflux environment is also strongly basic, distinguishing it from hydrothermal techniques, and operates at temperatures lower than those of hydroxide fluxes. This opens up novel phase spaces[16] and makes hydroflux synthesis a unique and useful tool for furthering materials discovery efforts.

Hydroflux synthesis involves heating a roughly equimolar solution of water and alkali hydroxide in a sealed reaction vessel at moderate temperatures. Here, $H_2O$ and AOH autodissociate into hydroxide ($[OH]^-$) and hydronium ($H_3O^+$) or alkali ($A^+$) species. These species are in dynamic equilibrium with other reactants, such as dissolved $O_2$ and $O^{2-}$,[17] and can form temperature- and concentration-dependent complexes with other introduced reagents, leading to the precipitation of new metastable phases.[18,19] These $O_2$ and $O^{2-}$ solution species, which can be stabilized by a high concentration of alkali hydroxide, influence the oxidation of the Te complexes and thus the structure of the phases formed.[20] Oxygen availability can be altered within the solution environment using $H_2O_2$ in place of $H_2O$,[17] further highlighting the versatility of the hydroflux technique.

Our previous work synthesizing complex layered tellurium oxides from a KOH-based hydroflux showed trends in structural dimensionality, protonation of oxygen, and oxidation of tellurium as functions of the concentrations of the reagents.[4] This motivated us to explore Cs-containing hydrofluxes in search of novel magnetic layered phases containing $Cs^+$, which could serve as a larger interlayer spacer relative to $K^+$. The adjustment of the distances between layers and the charge distribution within these layered phases is known to alter the electronic properties through confinement effects, changes to the coordination environment, and shifting electron counts.[9,21,22]

We chose to study Cu–Te–O systems for their potential to host novel magnetism.[23] Fully oxidized $Cu^{2+}$ has a $d^9$ outer shell and can act as a model spin $\frac{1}{2}$ ion. When ordered in a crystalline lattice, individual $Cu^{2+}$ ions can interact to form complex magnetic states based on their distances and geometries relative to each other and to nearby non-magnetic species that may facilitate superexchange. Tellurium was chosen because of its moderate solubility under hydroflux conditions and common non-magnetic oxidation states, $Te^{4+}$ and $Te^{6+}$.[24,25] The ions facilitate superexchange between magnetic $Cu^{2+}$ ions in different ways, since $Te^{6+}$ has an empty 5s orbital and coordinates octahedrally to oxygen, while $Te^{4+}$ has a stereochemically active 5s lone pair resulting in anisotropic coordination with oxygen.[20] Hydroflux synthesis of complex tellurates is well-reported,[26] but further studies are required to develop control over ligand identities and redox chemistry.

[a]Department of Physics, Oregon State University, 1500 SW. Jefferson Way, Corvallis, OR 97331, USA
[b]Department of Chemistry, Johns Hopkins University, 3400 N. Charles Street, Baltimore, MD 21218, USA. E-mail: aiwanic1@jhu.edu
[c]Institute for Quantum Matter, William H. Miller III Department of Physics and Astronomy, Johns Hopkins University, 3400 N. Charles Street, Baltimore, MD 21218, USA
[d]Department of Materials Science and Engineering, Johns Hopkins University, 3400 N. Charles Street, Baltimore, MD 21218, USA

† These authors contributed equally to this work.

Here, we report the single crystal synthesis of three novel phases, $CsTeO_3(OH)$, $KCu_2Te_3O_8(OH)$, and $Cs_2Cu_3Te_2O_{10}$, out of hydroflux solution (Table 1). $CsTeO_3(OH)$ is nonmagnetic and a member of the $ATeO_3(OH)$ series (A = Li, Na, K). $KCu_2Te_3O_8(OH)$ is magnetically three-dimensional, and undergoes spin ordering/reorientation transitions. $Cs_2Cu_3Te_2O_{10}$ consists of 2D planes of $Cu^{2+}$ trimers and $Te^{6+}$ dimers separated by disordered $Cs^+$ layers. It has no long range magnetic order down to $T$ = 2 K. Our results demonstrate the versatility of hydrofluxes in stabilizing unusual and complex, magnetically active bonding topologies.

# II. Experimental methods

### A. Synthesis

Samples were synthesized *via* hydroflux reactions as follows: powder reagents CuO (Thermo Scientific, 99.995%) and $TeO_2$ (ACROS Organics, 99%+) were combined in the ratios 1 : 10 mmol, 0.5 : 10 mmol, or 0 : 10 mmol. 3 mL of 30%, 10%, or 0% (pure water) aqueous $H_2O_2$ solution (Fisher Chemical, 30%) were combined with alkali hydroxides $KOH \cdot xH_2O$ (Fisher Chemical, 86.6%) or $CsOH \cdot xH_2O$ (Sigma-Aldrich, 90.0%) in the ratios 5 : 1, 7 : 1, or 10 : 1. For the alkali hydroxide reagents, percentages represent the molar amount of hydroxide relative to water. Reagents were loaded into a 22 mL capacity Teflon-lined autoclave with $H_2O_2$ added last and dropwise to minimize sudden $O_2$ gas formation. The autoclaves were heated to 200 °C for 2 or 3 days in a low temperature oven and quenched to room temperature on the benchtop. The samples were rinsed with 18 mΩ deionized (DI) $H_2O$ and filtered with a vacuum funnel. Optical microscopy images depict the novel phases on 1 $mm^2$ graph paper.

Table 1 SCXRD data of the novel compounds synthesized in this work: $CsTeO_3(OH)$, $KCu_2Te_3O_8(OH)$, $Cs_2Cu_3Te_2O_{10}$

| Compound | $CsTeO_3(OH)$ | $KCu_2Te_3O_8(OH)$ | $Cs_2Cu_3Te_2O_{10}$ |
|---|---|---|---|
| $T$ (K) | 213(2) | 213(2) | 213(2) |
| Space group | $P\bar{1}$ (2) | $P2_1/c$ (14) | $C2/m$ (12) |
| Crystal system | Triclinic | Monoclinic | Monoclinic |
| $a$ (Å) | 5.1861(2) | 13.2352(8) | 5.6630(2) |
| $b$ (Å) | 7.2579(4) | 7.8537(4) | 14.6253(6) |
| $c$ (Å) | 11.3948(6) | 9.6019(6) | 7.3375(3) |
| $\alpha$ (°) | 86.535(5) | 90 | 90 |
| $\beta$ (°) | 89.903(4) | 109.739(7) | 100.550(4) |
| $\gamma$ (°) | 89.073(4) | 90 | 90 |
| Volume (Å$^3$) | 428.06(4) | 939.43(10) | 597.44(4) |
| $Z$ | 4 | 2 | 2 |
| $\lambda$, Mo Kα (Å) | 0.71073 | 0.71073 | 0.71073 |
| No. reflections collected | 16298 | 18216 | 9086 |
| No. independent reflections | 3228 | 2731 | 1166 |
| $\theta$ range (°) | 2.812 to 32.998 | 3.066 to 29.997 | 2.785 to 33.000 |
| Index ranges | $-7 \le h \le 7$<br>$-11 \le k \le 11$<br>$-17 \le l \le 17$ | $-18 \le h \le 18$<br>$-11 \le k \le 11$<br>$-13 \le l \le 13$ | $-8 \le h \le 8$<br>$-22 \le k \le 22$<br>$-11 \le l \le 11$ |
| $F(000)$ | 556 | 1220 | 762 |
| Goodness-of-fit on $F^2$ (ref. 28) | 1.031 | 1.044 | 1.231 |
| $R_1 wR_2$ [$I \ge 2\sigma(I)$] [28] | 0.0286, 0.0501 | 0.0301, 0.0557 | 0.0257, 0.0656 |
| $R_1 wR_2$ [all data] [28] | 0.0463, 0.0558 | 0.0448, 0.0605 | 0.0281, 0.0668 |
| Largest diff. peak/hole (Å$^{-3}$) | 1.283/−1.360 | 1.345/−1.082 | 1.609/−1.299 |

$CsTeO_3(OH)$ crystals formed as white needles in spherical aggregates, exclusively growing on top of a black secondary phase. Our optimized synthesis of $CsTeO_3(OH)$ requires Cu : Te = 1 : 10 mmol, 30% $H_2O_2$(aq.) : CsOH = 10 : 1, and a dwell time of 2 days. Formation of $CsTeO_3(OH)$ was found to be very sensitive to concentration of aqueous $H_2O_2$ and the ratio $H_2O_2$(aq.) : CsOH, as tuning these parameters resulted in no $CsTeO_3(OH)$ formation (Fig. S3, S6 and S7). In addition, synthetic conditions of Cu : Te = 0 : 10 mmol, 30% $H_2O_2$(aq.) : CsOH = 10 : 1, (*i.e.* excluding CuO as a reagent) yielded no solid product. A phase with the same stoichiometry was mentioned in a previous study,[27] but no structure or additional characterization was provided.

$KCu_2Te_3O_8(OH)$ crystallized in clusters of irregularly shaped teal crystals. Our optimized, phase-pure synthesis requires Cu : Te = 0.5 : 10 mmol, 0% $H_2O_2$(aq.) : KOH = 10 : 1, and a dwell time of 3 days. Larger crystals were grown in a non-phase-pure synthesis (Fig. S2). There are no known alkali analogs of this phase.

$Cs_2Cu_3Te_2O_{10}$ crystallized as thin green plates in orthogonal clusters. Our optimized synthesis of $Cs_2Cu_3Te_2O_{10}$ requires Cu : Te = 0.5 : 10 mmol, 0% $H_2O_2$(aq.) : CsOH = 7 : 1, and a dwell time of 2 days. Reducing Cu : Te yielded greater phase purity at the cost of smaller overall yield. There are no known alkali analogs of this phase.

### B. Characterization

Single crystal X-ray diffraction (SCXRD) measurements were performed using a SuperNova diffractometer (equipped with Atlas detector) with Mo Kα radiation ($\lambda$ = 0.71073 Å) under the program CrysAlisPro (version 1.171.42.49, Rigaku OD, 2020–2022). The same program was used to refine the cell dimensions and for data reduction. All reflection intensities were measured at $T$ = 213(2) K. The structure was solved with the program SHELXS-2018/3 and was refined in $F^2$ with SHELXL-2018/3.[28] Analytical numeric absorption corrections or numerical absorption correction based on Gaussian integration over a multifacetssed crysal model were performed using CrysAlisPro. The temperature of the data collection was controlled using the Cryojet system (Oxford Instruments). The structural, lattice, and isotropic displacement parameters for all novel phases can be found in Table S1. Due to the large electron count on Cs and Te, hydrogen positions could not be determined reliably from SCXRD.

In all cases, we performed SCXRD on isolated single crystals to determine their structure. Since these crystals were not always the only precipitate, we also performed pXRD to characterize the identity and quantity of secondary phases. Powder X-ray diffraction (pXRD) measurements were performed using a Bruker D8 Focus diffractometer equipped with a LynxEye detector using Cu Kα radiation ($\lambda$ = 1.5406 Å). Data was collected in the range $2\theta$ = 5–120° with a step size of 0.01599° and a step time of 2 seconds. pXRD was performed on powder representative of the entire product yield for each synthesis, in order to accurately compare ratios of primary and secondary phases between optimization attempts. To do so, the entire sample was ground in an agate mortar and pestle after optical images were taken and mass was recorded. pXRD Rietveld refinements were performed using Topas5 using the refined single crystal structure as the starting point refinement for each compound.[29] Subsequently, lattice parameters and peak shape were refined and changed from the single crystal solution for all phases. For refinements of $Cs_2Cu_3Te_2O_{10}$ in Fig. 7 and Fig. S3, thermal displacement parameters and positions of the mobile Cs atoms were also refined to account for the different temperature conditions of the SCXRD and pXRD measurements. For the refinement of $KCu_2Te_3O_8(OH)$ in Fig. 4 and $CsTeO_3(OH)$ in Fig. 2, thermal displacement parameters and positions of all atoms were refined to account for these temperature differences. These parameters were not refined for other data to avoid overfitting.

Temperature-dependent magnetic susceptibility data was collected on a Quantum Design Magnetic Property Measurement System (MPMS3) from $T$ = 2–300 K with an applied field of $\mu_0H$ = 0.1 T using FC (field cooling) and ZFC (zero field cooling) modes. Isothermal magnetization measurements were collected at $T$ = 2, 10, 50, and 300 K with a range of $\mu_0H$ = ±7 T. All magnetic ordering results were collected on mechanically separated single crystals from multiphase products. Scanning electron microscopy (SEM) and energy dispersive spectroscopy (EDS) were performed using a JEOL JSM-IT100 by mounting single crystals on carbon tape. All crystal structure visualizations were performed using VESTA.[30]

## III. Results and discussion

### A. $CsTeO_3(OH)$

$CsTeO_3(OH)$ is a newly discovered analog to existing Li, Na, and K phases.[31–33] Our optimized growth requires Cu : Te = 1 : 10 mmol, 30% $H_2O_2$(aq.) : CsOH = 10 : 1, and a dwell time of 2 days. Despite growth from a solution containing CuO, no Cu is present in the structure. Analogous synthesis without CuO reagent yielded no solid precipitate, indicating that CuO, either as a solid or dissolved species, plays a crucial role in stabilizing formation of this phase under hydroflux conditions. Despite the requirement of CuO in this synthesis, this phase

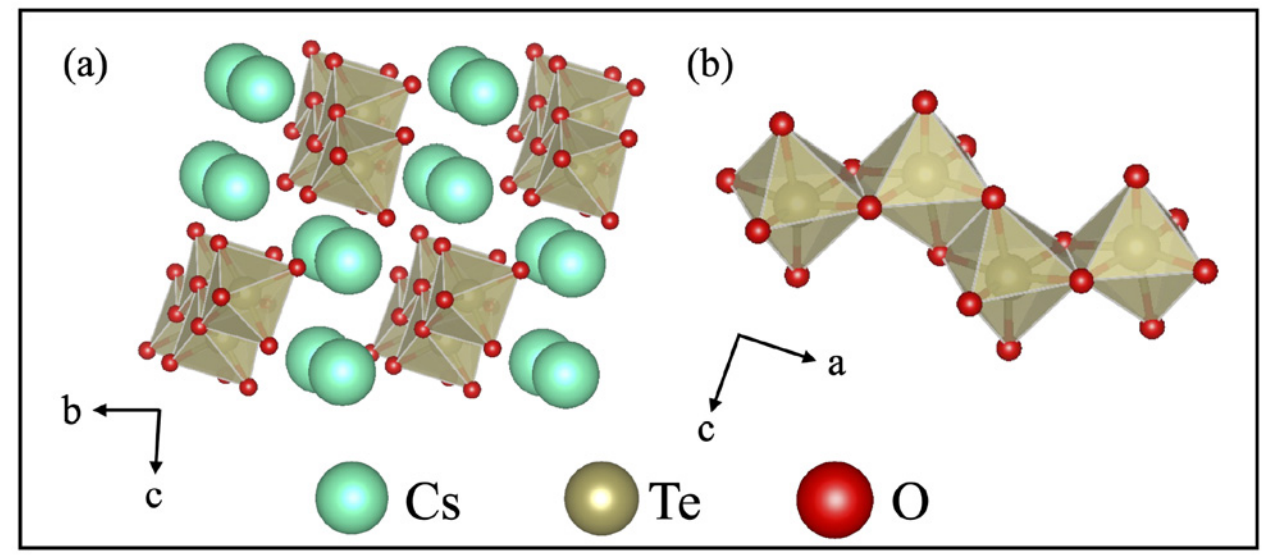

**Fig. 1** (a) The three-dimensional structure of $CsTeO_3(OH)$ has (b) chains of edge-sharing $TeO_6$ bridged by $Cs^+$ ions. Although the hydrogen positions cannot be resolved by SCXRD, it is likely that they occur between chains as in the K-analogue.[32]

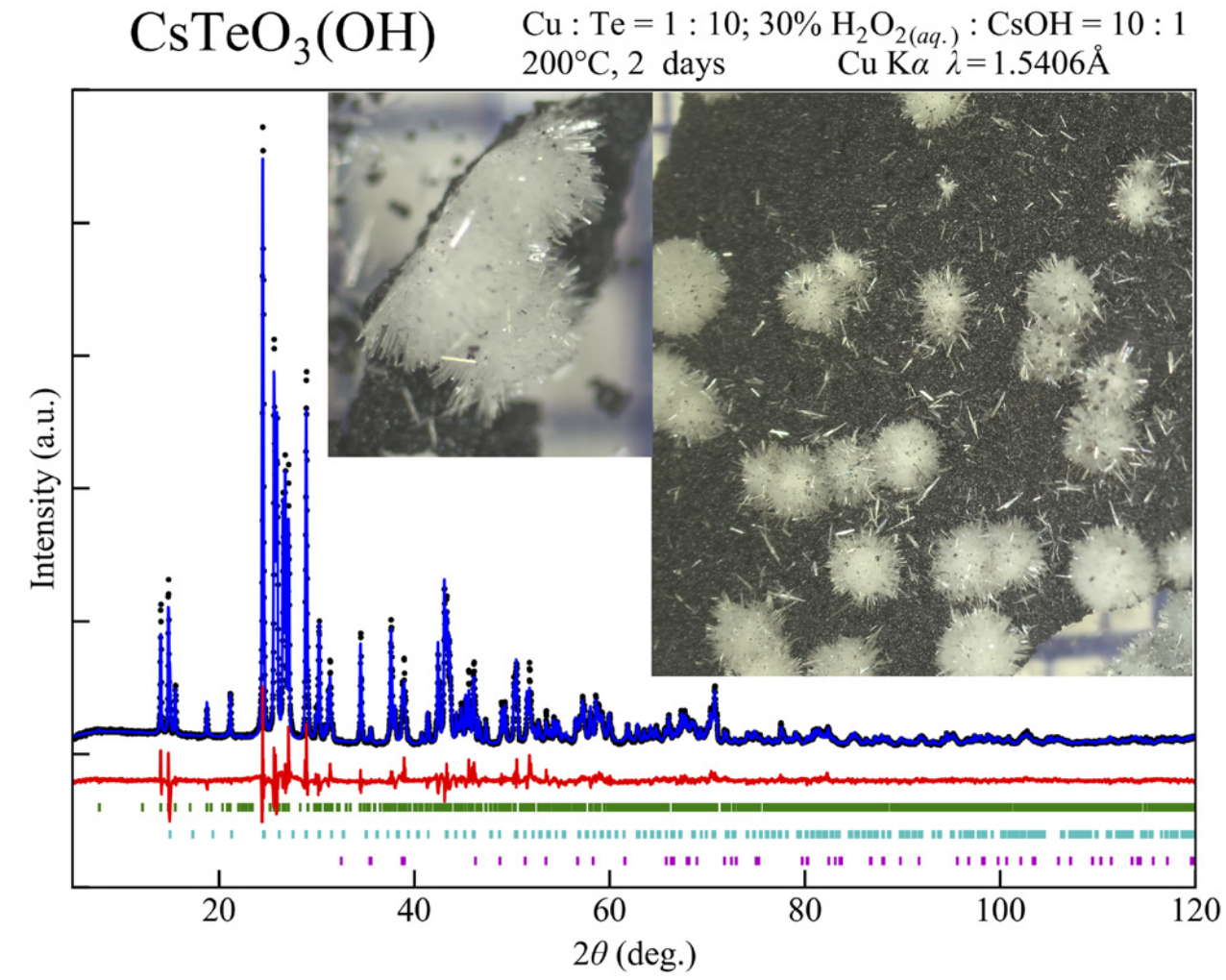

**Fig. 2** pXRD with Rietveld refinement ($R_{wp}$ = 10.77, $\chi^2$ = 3.27) and optical images of mixed-phase sample containing white $CsTeO_3(OH)$ crystals (green ticks, wt% Rietveld = 79.43(16)), $Cs_2Te_4O_{12-x}$ (cyan ticks, wt% Rietveld = 18.32(12)), and CuO (purple ticks, wt% Rietveld = 2.24 (14)). The oxygen content of the $Cs_2Te_4O_{12-x}$ powder was not refined.

precipitates exclusively on top of a powder defect pyrochlore $Cs_2Te_4O_{12-x}$, which does not contain copper. In addition, adjusting the ratio $H_2O_2$(aq.) : CsOH from 10 : 1 to 7 : 1, or adjusting the concentration of aqueous $H_2O_2$ from 30% to 10% or 0% resulted in entirely different phases (Fig. S4, S6 and S7).

$CsTeO_3(OH)$ is structurally three-dimensional with 9-fold coordinated $Cs^+$ and octahedrally coordinated $Te^{6+}$ ions (Fig. 1). Compared to the analogs $ATeO_3(OH)$ (A = Li, Na, K) that crystallize in monoclinic unit cells,[33] $CsTeO_3(OH)$ crystallizes in the lower-symmetry triclinic space group $P\bar{1}$, and with a larger unit cell than its analogs, reflective of its larger ionic radius. A visual comparison between $KTeO_3(OH)$ and $CsTeO_3(OH)$ can be found in Fig. S8. Distorted $TeO_6$ octahedra, with O–Te–O bond angles varying between 78.15° and 96.59°, form edge-sharing chains in $CsTeO_3(OH)$. While the hydrogen positions could not be reliably resolved using SCXRD, a poss-ible hydrogen position is shown in the Fig. S8. To definitively determine the position, additional evidence from *e.g.* neutron diffraction studies would be useful. Unlike the other novel phases presented in this paper, which contain copper, $CsTeO_3(OH)$ is diamagnetic due to its lack of unpaired spins.

This phase may be useful as a reactive precursor to novel tellurate phases because of the presence of hydroxide ligands, which are expected to undergo a dehydration reaction at moderate temperatures accompanied by structural transformations which may be conducive to the synthesis of other complex oxides.

### B. $KCu_2Te_3O_8(OH)$

Single crystals of $KCu_2Te_3O_8(OH)$ formed phase-pure in a synthesis with Cu : Te = 0.5 : 10 mmol, 0% $H_2O_2$(aq.) : KOH = 10 : 1, and a dwell time of 3 days. When other parameters were held constant and Cu : Te = 1 : 10 mmol, larger crystals formed but with lower phase purity (Table S3 and Fig. S2). The hydroflux reaction environment with 0% $H_2O_2$(aq.) : KOH = 10 : 1 was not sufficiently oxidizing to oxidize the $Te^{4+}$ reagent to $Te^{6+}$.

This phase (Fig. 3) contains $Te^{4+}$ in an anisotropic coordination due to its stereochemically active lone pair, which tends to orient towards interstitial spaces. The Te–O bond lengths and angles vary. $Cu^{2+}$ takes a square planar coordination, with additional apical Cu–O interactions of varied strength. These apical interactions are not expected to be strongly involved in magnetic pathways due to poor orbital overlap and long bond length. Due to the presence of heavy tellurium, we were unable to determine the exact location of the hydrogen atom from our SCXRD data.

Magnetic susceptibility of this phase was probed as a function of field and temperature. As shown in Fig. 5, two magnetic features are observed: one broad, short-range antiferromagnetic transition at $T$ = 21.7 K and a ferromagnetic correlation at $T$ = 29.2 K. These transition temperatures were determined with a derivative analysis, as seen in Fig. S9. Given the complex coordination environments of both Te and Cu, the precise determination of magnetic pathways is challenging and requires further investigation through density functional theory (DFT) modeling, similar to those performed in ref. 34, or neutron diffraction studies.[35] Isothermal magnetization measurements suggest the presence of possible metamagnetic transitions at $\mu_0 H \approx 2$ T at both $T$ = 2 K and $T$ = 10 K.

A Curie–Weiss fitting was performed in the paramagnetic region above $T$ = 150 K:

$$\chi_{\text{mol}} = \chi_0 + \frac{C}{T - \theta_{\text{CW}}} \tag{1}$$

where $C$ is the Curie constant which is related to the effective magnetic moment ($\mu_{\text{eff}} = \sqrt{(8C)}$) and $\theta_{\text{CW}}$ is the Weiss temperature, which gives an indication of net interactions of the system. From this fitting, we extracted $\theta_{\text{CW}} = -138.6 \pm 0.2$ K, $C = 0.5863 \pm 0.0005$ emu K mol$^{-1}$, and $\mu_{\text{eff}} = 2.16\mu_B$. This is consistent with the antiferromagnetic transitions observed in this phase; however, the effective moment is higher than the spin-only moment expected from an isolated spin $\frac{1}{2}$ ion. The theoretical spin-only magnetic moment is given by

$$\mu = \sqrt{N(N+2)}\mu_B \tag{2}$$

where $N$ is the number of unpaired electrons and $\mu_B$ is the Bohr magneton. For $KCu_2Te_3O_8(OH)$, there is one unpaired electron ($N$ = 1) per Cu site, resulting in a theoretical spin-only moment of $\mu = 1.73\mu_B$.

As highlighted above, the magnetic properties of $Cu^{2+}$-containing crystals are sensitive to the coordination of the constituent atoms. Structure-dependent magnetic behavior has been observed in other square planar $Cu^{2+}$ compounds

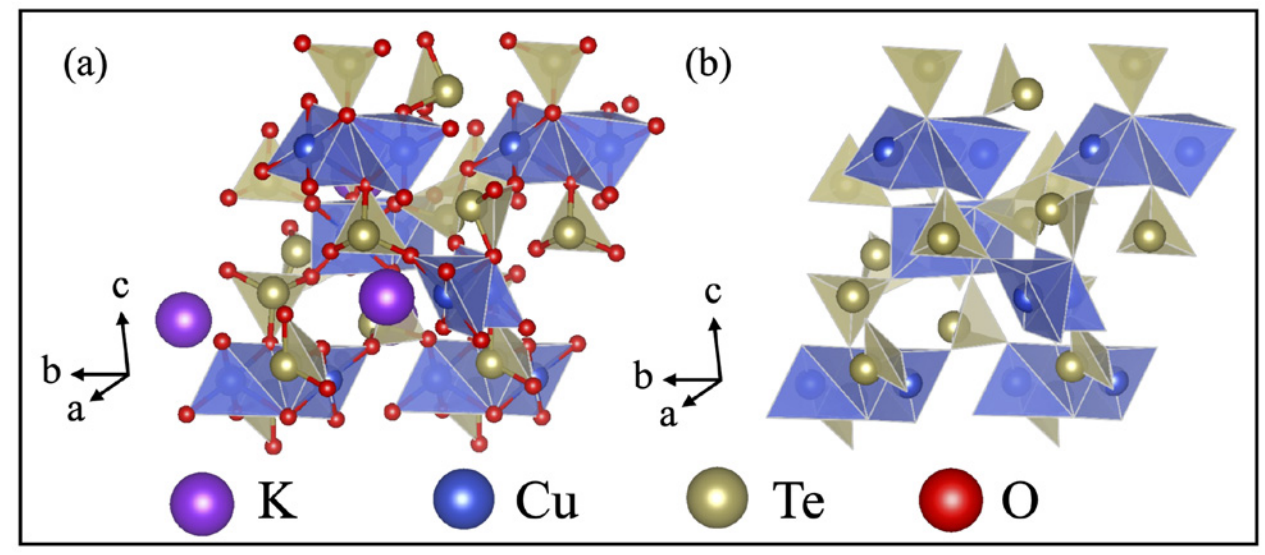

Fig. 3 (a) The three-dimensional structure of $KCu_2Te_3O_8(OH)$ and (b) the Cu–Te substructure. Note the anisotropic $Te^{4+}$ coordination due to the lone pair effect. Position of the hydrogen atom has not been resolved, and no alkali analogs are known for this phase.

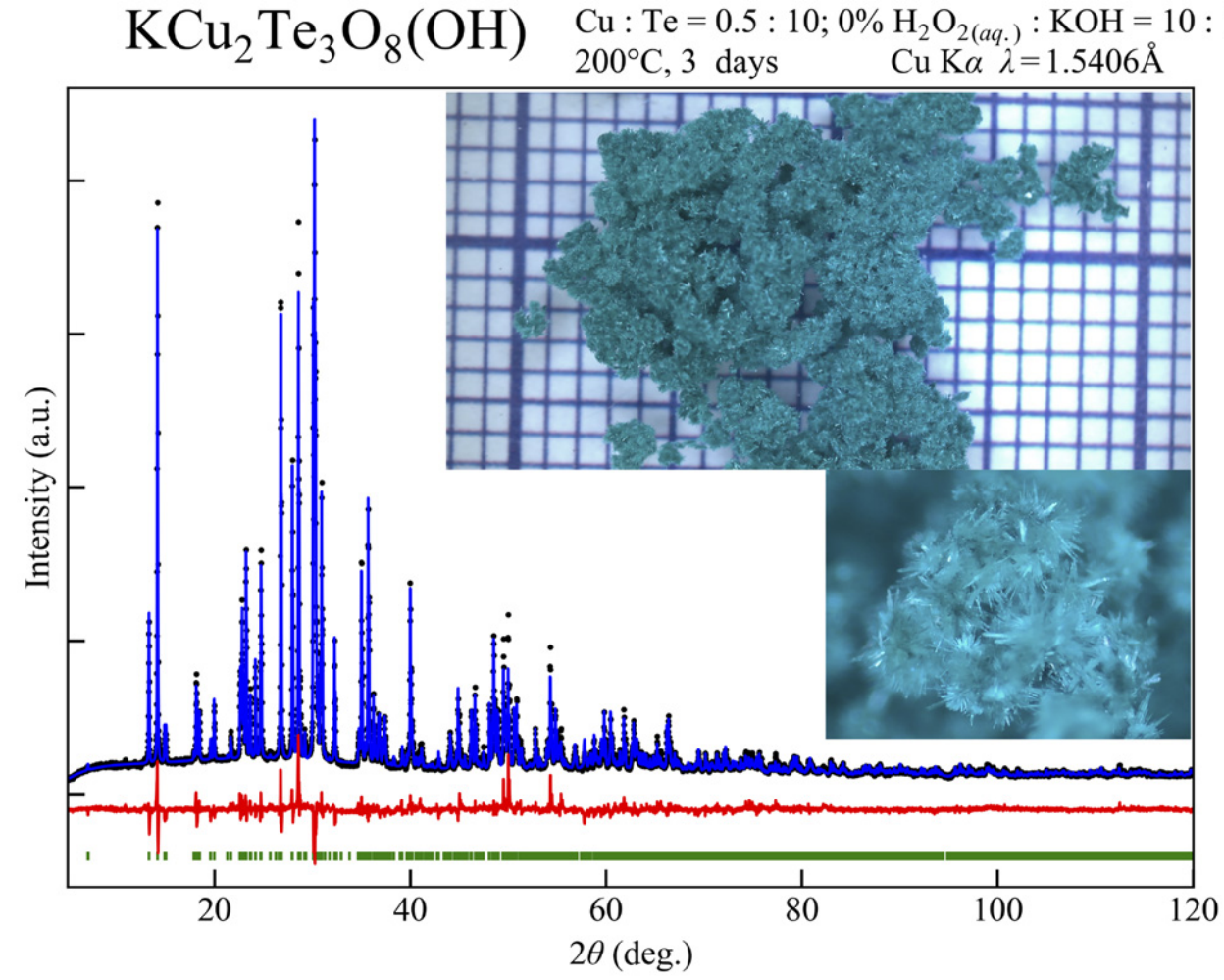

Fig. 4 pXRD with Rietveld refinement ($R_{wp}$ = 7.55, $\chi^2$ = 2.58) and optical images of phase-pure $KCu_2Te_3O_8(OH)$ (green ticks). Atomic positions and thermal displacement parameters were refined to account for temperature differences between SCXRD and pXRD measurement conditions. These refined positions resulted in the same atomic connectivity with some rotations as would be expected from slight additional thermal energy.

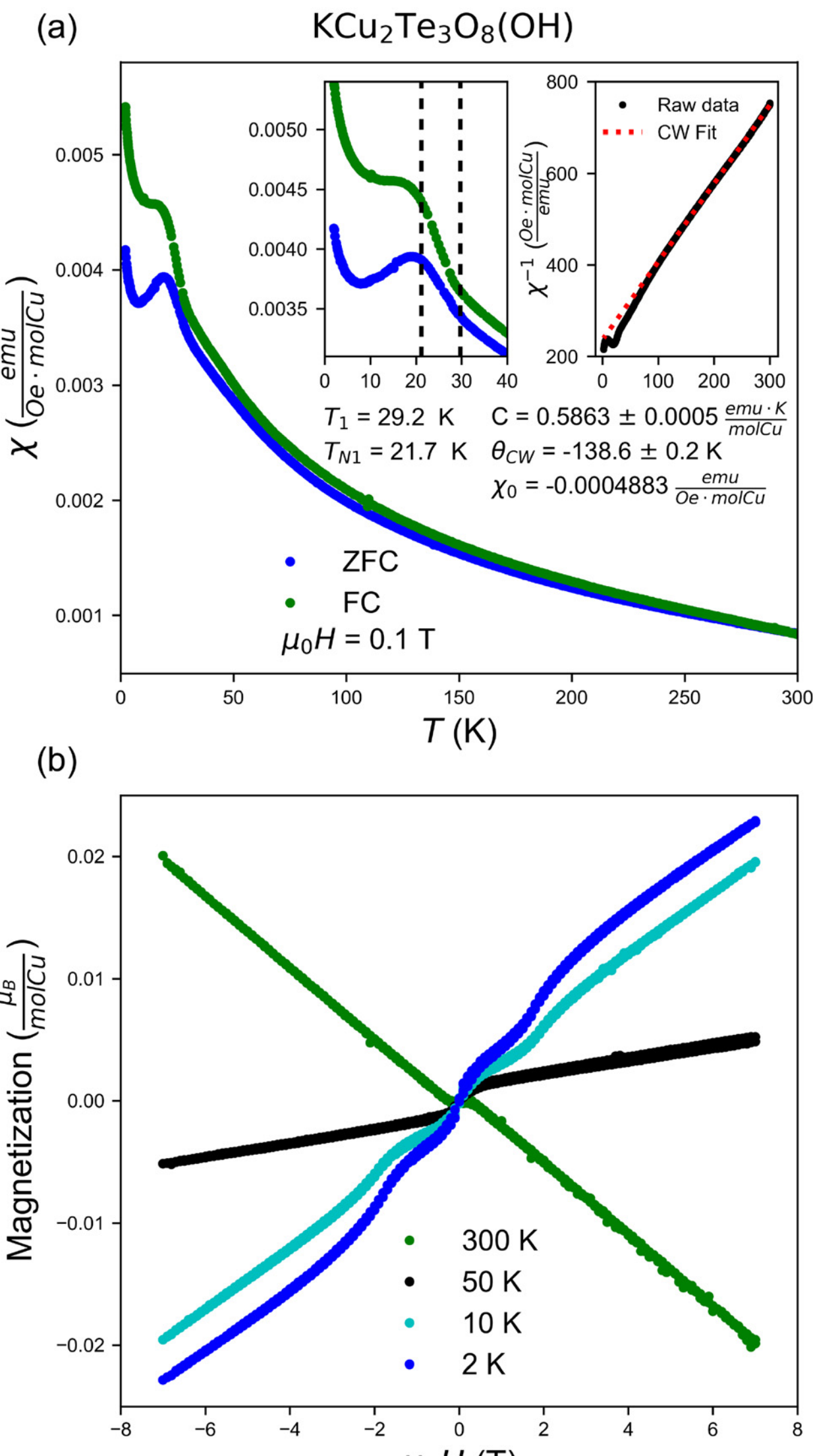

**Fig. 5** (a) Magnetic susceptibility measurements of $KCu_2Te_3O_8(OH)$ were performed under ZFC and FC conditions with an applied external field of $\mu_0H$ = 0.1 T. The data reveal two distinct magnetic features: one short-range antiferromagnetic transition at $T$ = 29.2 K, and a ferromagnetic correlation at $T$ = 21.7 K. Insets show the low temperature transitions and Curie–Weiss fit, with $\theta_{CW}$ = −138.6 ± 0.2 K. (b) Isothermal magnetization data for $KCu_2Te_3O_8(OH)$ were collected at temperatures of $T$ = 2, 10, 50, and 300 K, within the external magnetic field range of $\mu_0H$ = ±7 T. The magnetization curves at $T$ = 2 K and $T$ = 10 K suggest the presence of possible metamagnetic transitions occurring at $\mu_0H \approx 2$ T.

recently reported by our group,[4] as well as in tetrahedral $Cu^{2+}$ in $Cu_2Te_2O_5X_2$ (X = Cl, Br), as described in ref. 36. The varied degree of orbital overlap in these compounds gives rise to a range of behavior; to fully understand the diversity of magnetic interactions in $Cu^{2+}$ compounds and their dependence on local structure, new phases must be synthesized and their magnetic superexchange mechanisms must be computationally analyzed.

### C. $Cs_2Cu_3Te_2O_{10}$

Our optimized conditions for single crystal growth of $Cs_2Cu_3Te_2O_{10}$ require Cu : Te = 0.5 : 10 mmol, 0% $H_2O_2$(aq.) : CsOH = 7 : 1, and a dwell time of 2 days. Changing Cu : Te, $H_2O_2$(aq.) : CsOH, and aqueous peroxide concentration were found to decrease phase purity and yield (Table S3 and Fig. S3–S5). These hydroflux conditions (some without peroxide) were sufficiently oxidizing to form this $Te^{6+}$ phase from the $Te^{4+}$ reagent. Attempts to synthesize a K-analog from a KOH-based hydroflux were unsuccessful, but syntheses of analogs using alkali (de)intercalation methods remain of interest to tailor interlayer magnetic exchange pathways.

$Cs_2Cu_3Te_2O_{10}$ contains alternating layers of disordered $Cs^+$ ions and ordered Cu–Te–O planes of $TeO_6$ octahedra and $CuO_4$ square-planar plaquettes. The disorder in the Cs layer suggests possible ion mobility at and above the SCXRD measuring temperature of $T$ = 213 K. The Cu–Te–O layer is structurally two-dimensional, but contains significant bonding along the $b$ direction, with three edge-sharing $CuO_4$ 'trimers' between two edge-sharing $TeO_6$ 'dimers' as can be seen in Fig. 6c. The three $CuO_4$ square-planar plaquettes display similar Cu–O bond lengths and O–Cu–O bond angles. However, significant differences are observed in the orientation of these bonds, such that the one central and two outer Cu within a trimer are in distinct coordination environments. Specifically, within a group of three edge-sharing $CuO_4$ plaquettes, the central plaquette contains all Cu–O bonds in a single plane while the outer plaquettes have Cu–O bonds twisted out of a single plane. The Cu trimers and Te dimers are offset along the $a$ direction such that the 'planar' $CuO_4$ plaquette bonds along the $a$ direction apically to $TeO_6$. Since planar and twisted plaquettes share oxygen atoms, the 'twisted' $CuO_4$ plaquettes also bond to Te apically along the $a$ direction, as well as doubly along the $b$ direction. These bonding pathways facilitate magnetic superexchange in two dimensions.

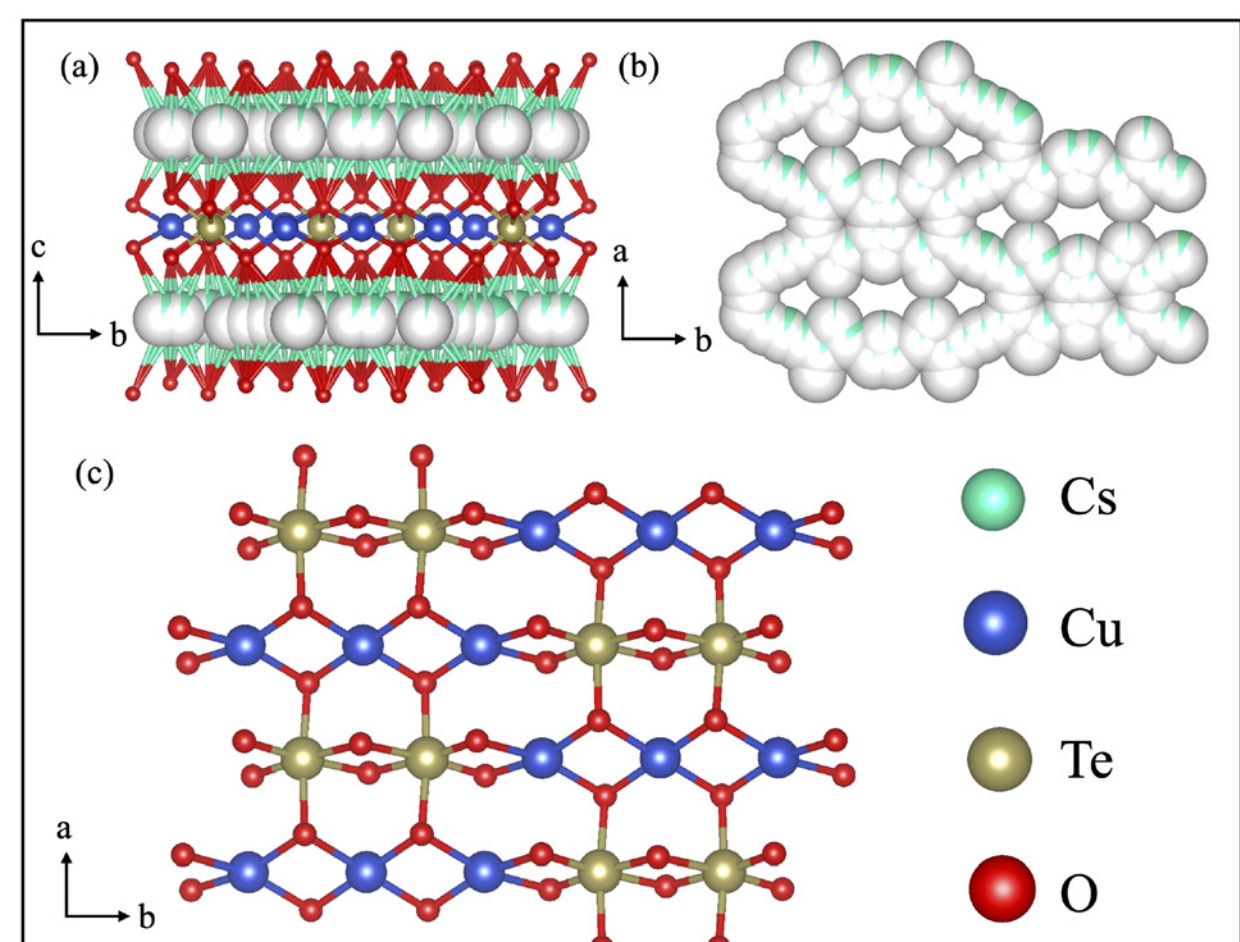

**Fig. 6** Structure of layered $Cs_2Cu_3Te_2O_{10}$ showing (a) the layer stacking sequence (b) the disordered Cs layer, with cyan slivers in white spheres indicating partial occupancy of Cs and (c) the two-dimensional Cu–Te–O layer.

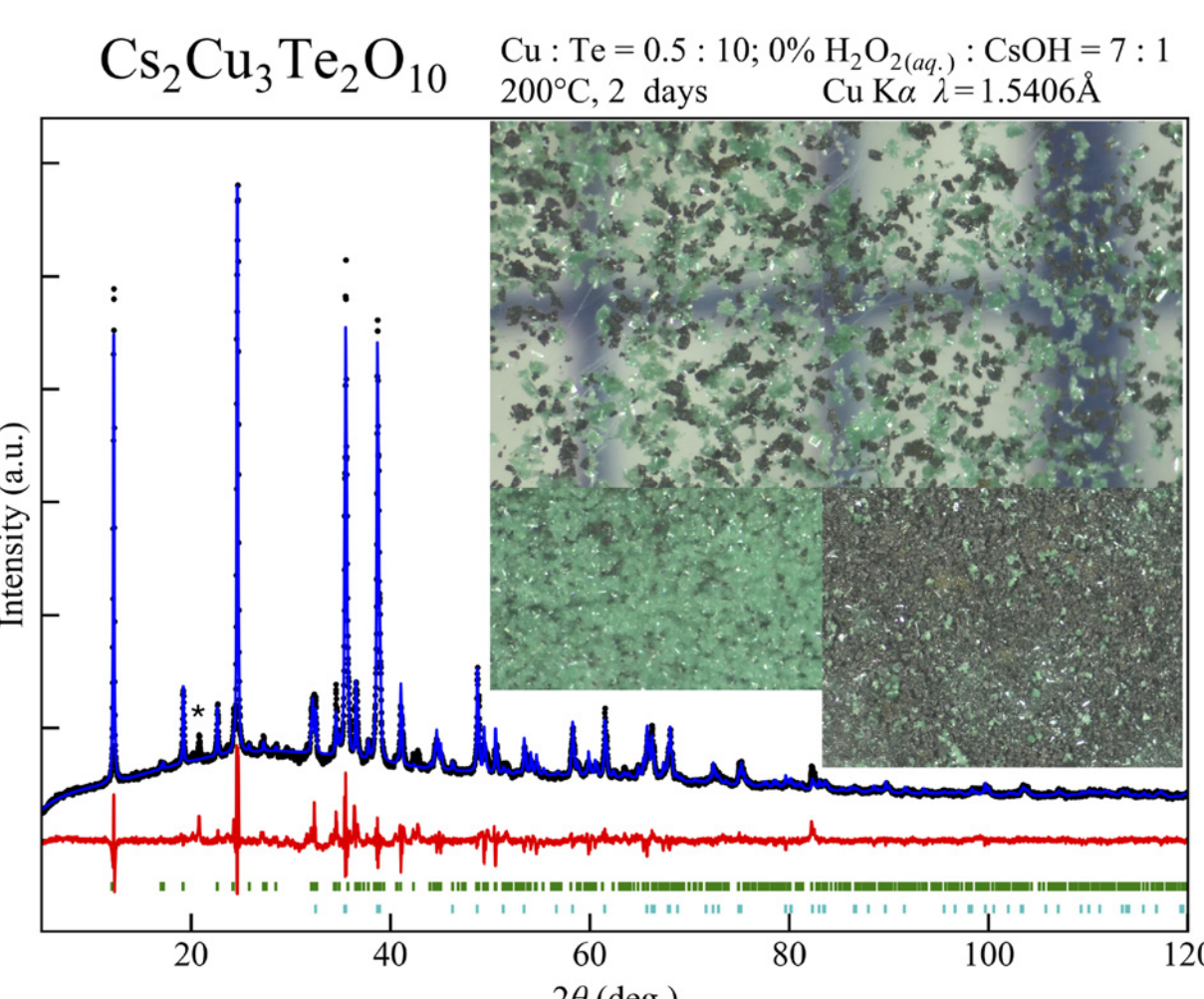

**Fig. 7** pXRD with Rietveld refinement ($R_{wp}$ = 5.14, $\chi^2$ = 2.86) and optical images of mixed-phase sample containing green $Cs_2Cu_3Te_2O_{10}$ crystals (green ticks, wt% Rietveld = 32.6(3)) and black CuO (cyan ticks, wt% Rietveld = 67.4(3)). Atomic positions and thermal displacement parameters of the Cs ions were refined to account for temperature differences between SCXRD and pXRD measurement conditions. The Cs ion positions changed slightly in-plane as would be expected from slight additional thermal energy. Asterisks correspond to residual peaks.

A disordered, 7 Å thick layer of Cs ions separates the Cu–Te–O layers, preventing magnetic superexchange between layers. The stoichiometry of Cs was determined using SCXRD and confirmed with EDS measurements as detailed in Table S2. Two $Cs^+$ ions per formula unit would charge-balance the Cu–Te–O layers.

The ZFC and FC $\chi$ *vs.* $T$ measurements reveal paramagnetic behavior, as shown in Fig. 8, which is corroborated by the isothermal magnetization data. Curie–Weiss analysis suggests net ferromagnetic interactions with $\theta_{CW}$ = 2.55 ± 0.1 K, $C$ = 0.2637 ± 0.0001 emu K mol$^{-1}$, and $\mu_{eff}$ = 1.45$\mu_B$. This calculated moment is slightly lower than the spin-only moment expected from an isolated spin-$\frac{1}{2}$ ion, $\mu$ = 1.73$\mu_B$.

The absence of ordering in this phase above $T$ = 2 K is likely related to several factors. First, large distances between layers prevent significant interlayer magnetic interaction. Then, within a single layer, interactions along the $a$ direction are expected to be weak, despite bonding in this direction, due to poor directional overlap between the Te orbitals and the Cu $d_{x^2-y^2}$ orbitals which are involved in magnetic exchange. In the $b$ direction, the twisting of the $CuO_4$ plaquettes reduces the overlap of adjacent $d_{x^2-y^2}$ orbitals. Thus, magnetic interactions are weak in three, two, or one dimension. Further experimental investigations are necessary at lower temperatures to determine the nature of any possible ordering in this phase.

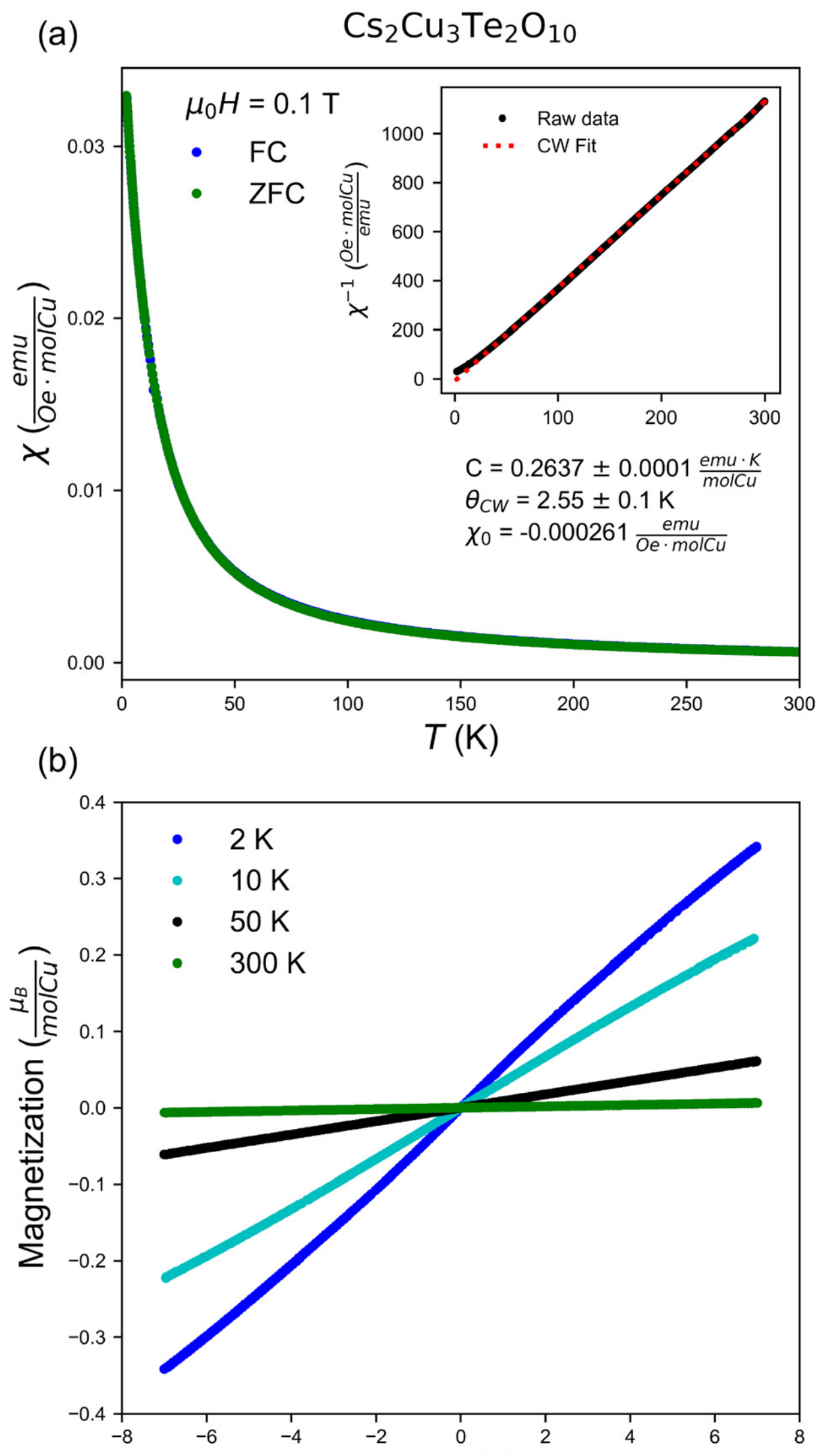

**Fig. 8** (a) Magnetic susceptibility measurements of $Cs_2Cu_3Te_2O_{10}$ were performed under ZFC and FC conditions with an applied external field of $\mu_0H$ = 0.1 T. Inverse magnetization (inset) shows Curie–Weiss-like behavior with $\theta_{CW}$ = 2.55 ± 0.1 K, indicating net ferromagnetic interactions. (b) Isothermal magnetization data for $Cs_2Cu_3Te_2O_{10}$ were collected at temperatures of $T$ = 2, 10, 50, and 300 K, within the external magnetic field range of $\mu_0H$ = ±7 T.

## D. Formation trends

All phases obtained in this study precipitated from a hydroflux solution with a CuO : $TeO_2$ molar ratio of 1 : 10 or 0.5 : 10. Preliminary reactions using CsOH-hydrofluxes with CuO : $TeO_2$ = 1 : 1 showed poor solubility of CuO. In contrast, in the KOH phase space, CuO : $TeO_2$ = 1 : 1 was well-solubilized.[4] Studies of alkali flux reactions have shown that the solubility of CuO in alkali hydroxide decreases as the size of alkali increases (that is, solubility in LiOH > NaOH > KOH).[37] Our hydroflux reactions in CsOH and KOH are consistent with this result. Despite the low CuO : $TeO_2$ molar ratio of reagents used in this work, the novel phases containing both Cu and Te had comparable (Cu : Te = 3 : 2 or 2 : 3) incorporation of these elements.

While we do not know the identity of Te complexes in solution, these observations highlight the markedly higher solubi-lity of Te-containing species compared to Cu species under these hydroflux conditions. Additional studies are required to determine the formation pathway responsible for the creation of the novel phases reported.

We can also compare the formation conditions of these novel phases to similar phases from our previous study.[4] In that study, increasing the concentration of hydroxides relative to water tended to decrease the protonation of oxygens within the structure, from structures containing water molecules to hydroxides to oxides. Herein, we found that formation of the novel oxide phase, $Cs_2Cu_3Te_2O_{10}$, occurred under conditions such that $H_2O_2$(aq.) : CsOH = 7 : 1 and 5 : 1, while formation of the novel oxide-hydroxide phases, $CsTeO_3(OH)$ and $KCu_2Te_3O_8(OH)$, occurred in environments with $H_2O_2$(aq.) : CsOH = 10 : 1 and $H_2O_2$(aq.) : KOH = 10 : 1 respectively. This trend has been seen elsewhere in the literature;[5] further study of hydroflux syntheses with dilute hydroxide may result in additional novel mixed oxide-hydroxide phases. As expected, the use of peroxide solution as oxidizer resulted in oxidized $Cu^{2+}$–$Te^{6+}$ phases; the only phase with partially oxidized $Te^{4+}$ formed in 0% $H_2O_2$ solution. We do not know the oxidation state of Te in the secondary phase $Cs_2Te_4O_{12-x}$ since the Cs and O content were not determined from the pXRD. However, under sufficient hydroxide concentration, peroxide is not necessarily required to form $Te^{6+}$, as can be seen in our $Cs_2Cu_3Te_2O_{10}$ syntheses.

## IV. Conclusions

In this study, we identified three novel phases within the tellurium oxide-hydroxide phase space: $CsTeO_3(OH)$, $KCu_2Te_3O_8(OH)$, and $Cs_2Cu_3Te_2O_{10}$. $CsTeO_3(OH)$ is a member of the nonmagnetic series of alkali tellurate oxide hydroxides $ATeO_3(OH)$ (A = Li, Na, K), with edge-sharing chains of $TeO_6$ octahedra. $KCu_2Te_3O_8(OH)$ is structurally three dimensional and contains three-coordinate $Te^{4+}$ ions and apical-oxygen-sharing $Cu_2O_8$ dimers with complex temperature- and field-dependent magnetic ordering. $Cs_2Cu_3Te_2O_{10}$ is a layered phase with Cu–Te–O layers separated by disordered Cs ions and no magnetic ordering above $T$ = 2 K.

Beyond the individual phases synthesized here, this study has highlighted valuable insights into hydroflux synthesis. The chemistry of these systems is rich and underexplored. Further investigations into other mixed alkali hydroxide flux systems, beyond the materials studied here, are likely to lead to the discovery of additional novel phases.

## Conflicts of interest

There are no conflicts to declare.

## Data availability

Data sets generated during the current study are available at https://doi.org/10.34863/mjqe-q055.

Supplementary information (SI) is available. See DOI: https://doi.org/10.1039/d5dt01468a.

CCDC 2466252–2466254 contain the supplementary crystallographic data for this paper.[38a–c]

## Acknowledgements

The MPMS3 system used for magnetic characterization was funded by the National Science Foundation, Division of Materials Research, Major Research Instrumentation Program, under Grant #1828490. This work made use of the synthesis facilities of the Platform for the Accelerated Realization, Analysis, and Discovery of Interface Materials (PARADIM), which is supported by the National Science Foundation under Cooperative Agreement No. DMR-2039380. M.G. acknowledges the REU-Site: Summer Research Program at PARADIM through grant number DMR-2150446.